# Design and performance of an absolute gas refractometer based on quasi-synthetic wavelength method

Jitao Zhang, Pei Huang, Yan Li and Haoyun Wei
*State Key Laboratory of Precision Measurement Technology and Instruments, Department of Precision Instruments and Mechanology, Tsinghua University, Beijing 100084, China*
zhangjt06@gmail.com

We present a refractometer which is capable of measuring the refractive index of gas with an unambiguous range of 1.000395 and uncertainty of 3.1E-8 at 633 nm absolutely. The measurement range is extended by means of a group of vacuum tubes according to the principle of quasi-synthetic wavelength (QSW) method. The basic principle of the QSW method and the design of the gas refractometer are presented in detail. The performance of the refractometer has been verified by the measurement of dry air, nitrogen gas and ambient air under different environmental situations. The gas-filling or pumping process is not needed during the measurement, so that we can complete a measurement within 70 seconds. Comparing with previous ones, the refractometer reported here has integrated virtues of large unambiguous range, fast speed, high accuracy, and a simple instrumentation design.

## 1 Introduction

The refractive index of ambient gas is usually a crucial data for optical instruments with high performance. For instance, most of the optical interferometers are performed in atmosphere, and the accurate knowledge of the refractive index of air is tightly related to the precision of length measurement. In addition, as the major constituent of air, nitrogen gas at normal condition is inert for chemical reaction and transparent at wavelengths in ultraviolet band. These properties make it an attractive purge gas used for deep ultra-violet lithography. Since high resolution imaging is needed, the refractive index of purge gas should be measured accurately for the optical design of photolithography system. The refractive index of air has been measured by many methods and currently can be estimated by empirical equations with an uncertainty of a few parts in $10^8$[1-3]. However, the empirical equation is only valid to standard air with specific proportion of gas components, so that its accuracy is usually questionable because of the variation of the components of practical ambient air. Therefore, the direct measurements are usually preferred in place where high accuracy or in-situ monitoring is required. The gas refractometer for direct measurement often derives the refractive index of a sample gas from comparing with that of a known standard, namely, vacuum. The direct methods can be categorized by two general approaches according to the measurement range. The first approach is called relative measurements, since they can only measure the relative fluctuation of the refractive index with their limited measurement ranges [4-5]. In other words, the initial value obtained by other methods is necessary to deduce the absolute value of the refractive index for the relative measurements. This approach limits its application to the known gases. The second approach is called absolute measurements, and their measurement ranges can cover the absolute value of measured gases. To expand the measurement range, one effective method is to record the refractivity data continuously corresponding with the process of gas-pumping from ambient to vacuum and vice versa [6-8]. Technically speaking, the measurement range can be as large as one's desire in this way. However, the gas-pumping process makes this method time-consuming and complex. Moreover, the fluctuation of gas pressure and temperature cannot be avoided in the measurement process. The idea of synthetic wavelength or exact fractions in the optical interferometry was also tried for measuring gas's refractive index absolutely[9]. Unfortunately, the lack of proper candidates of monochromatic light sources makes it difficult to obtain an expected measurement range. Recently, several absolute gas refractometers have been developed by means of Fabry-Perot cavity, optical frequency comb, and trapezoidal cavity[10-13]. For the Fabry-Perot cavity[10] and frequency combs routes[11-12], the systems are so sophisticated and expensive that they can be only used in laboratory. Also, additional measurement in vacuum environment is essential to deduce the final result. And for the trapezoidal cavity route[13], a specific trapezoidal cavity should be fabricated precisely. In this paper, we develop a gas refractometer which can measure gas refractivity absolutely and directly based on the quasi-synthetic wavelength (QSW) method. In this method, different from synthetic wavelength theory[14-15], the "synthetic wavelengths" are generated by a group of home-made vacuum tubes but not monochromatic light sources. The measurement range of this refractometer can be customized according to the measured gases. In addition, the gas-filling or pumping process is not necessary during the measurement, so that a complete measurement can be finished within 70 second. The simple configuration makes it compact and easy to handle. In the rest of the paragraph, we first explain the measurement principle of QSW method and the theoretical design of the refractometer in Section2. In Section 3, we construct the gas refractometer and verify its performance by measuring the dry air, nitrogen gas and ambient air directly. The discussion and conclusion are presented finally.

## 2 Principles and Method

### *2.1 QSW Method*

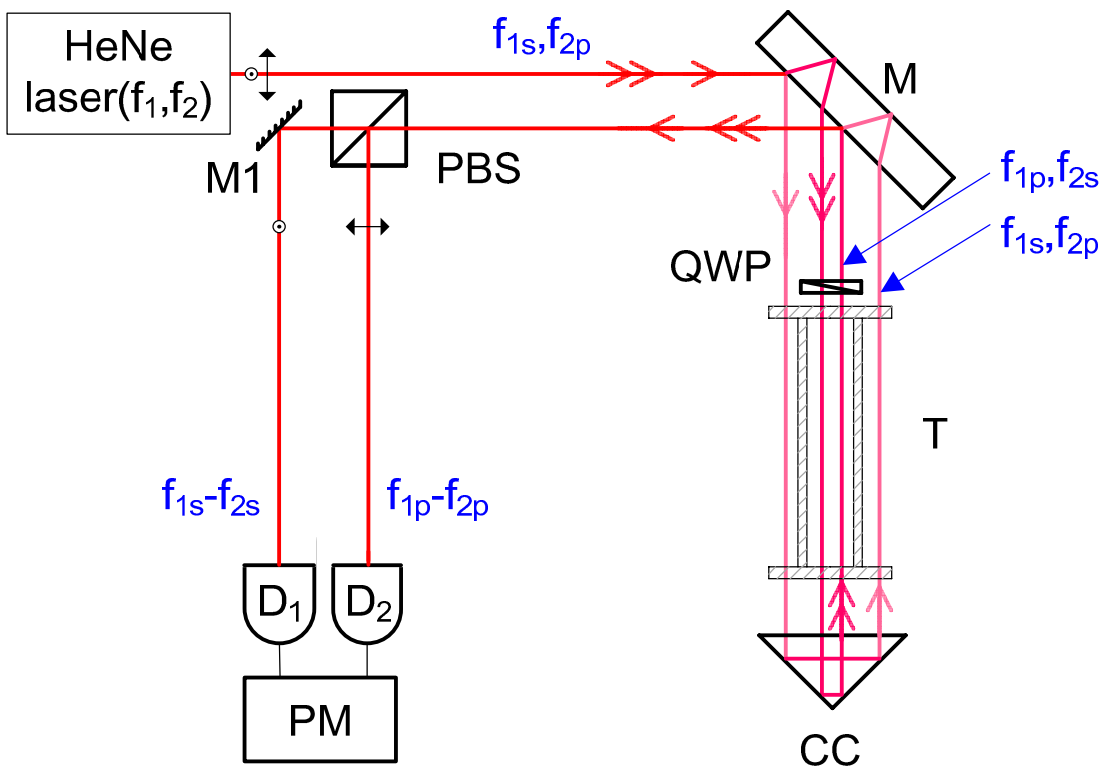

**Fig.1** Measurement principle of the gas refractometer. HeNe laser, laser with two orthogonally linear polarized frequency components f1 and f2, the subscript p and s indicates that the polarization direction of f1 and f2 is parallel and perpendicular to the printing paper, respectively;

M, plane beam splitter; QWP, quarter wave plate; T, vacuum tube; CC, corner cube; PBS, polarized beam splitter; M1, reflector; D1, D2, photo detectors; PM, phase meter.

The measurement principle of gas refractometer is shown in Fig.1.The polarization direction of two-frequency components f1 and f2 of the HeNe laser is represented by double-headed arrow and concentric circle in Fig. 1, respectively. The M is coated with 50% reflective film on the upper surface and >99% total reflection film on the bottom surface. When the incident angle is 45 degree, the HeNe laser can be divided by the M into two parallel beams with similar intensity (shown by double and single arrow in Fig. 1), which then passes through the inner and outer space of the vacuum tube T, respectively. After reflected by the CC, these two beam overlay at the upper surface of the M again and thus interfere. Then, the interfering beam is separated by the PBS into two different polarized components and detected by the detectors D1 and D2. A QWP is placed between the reflector M and the vacuum tube T, and the fast axis of which is adjusted as 45 degree to the polarization direction of f1- and f2-component coming from the laser source. The size of this QWP is selected properly so that only the inner light can pass through it. Owing to the QWP, the polarization direction of the inner light is rotated by 90 degree after twice-pass but that of the outer light remain the same. The air in the inner space of the vacuum tube has been exhausted by a vacuum pump in advance.

It can be deduced easily that the light received by D1 consists of f1s and f2s component passing through the outer and inner space of T, respectively. Similarly, the light received by D2 consists of f2p and f1p component passing through the outer and inner space of T, respectively. Therefore, the signal $I_1$ and $I_2$ detected by the detectors can be expressed as

$$\left.\begin{aligned} I_1 &= I_{01}\cos\left(2\pi\Delta ft + \Delta\varphi_0 + \Delta\varphi\right) \\ I_2 &= I_{02}\cos\left(2\pi\Delta ft + \Delta\varphi_0 - \Delta\varphi\right) \end{aligned}\right\}, (1)$$

where $I_{01}$ and $I_{02}$ are constant, $\Delta f = f_1 - f_2$ (suppose $f_1 > f_2$ ), $\Delta\varphi_0$ is the difference of initial phases of two frequency components, and $\Delta\varphi$ is the phase shift related to the difference of refractive index between ambient gas and vacuum. The opposite sign of $\Delta\varphi$ in the two formulas in Equ.1 is caused by different optical paths passed by two frequency components. The phase shift can be described as

$$\Delta\varphi = 2\pi\cdot\frac{2(n-1)L}{\lambda}, (2)$$

where $n$ is the refractive index of measured gas, $L$ is the length of T, and $\lambda$ is the wavelength of the laser source. Since the phase difference between D1 and D2 is twice of $\Delta\varphi$, and if we normalize the phase difference by $2\pi$ , then it can be expressed as

$$\frac{2\Delta\varphi}{2\pi} = N + \varepsilon = \frac{4(n-1)L}{\lambda}, (3)$$

where $N$ and $\varepsilon$ are integral and fractional part of the normalized phase. And Equ.3 can be rewritten as

$$n-1 = \lambda_s\left(N+\varepsilon\right), (4)$$

where $\lambda_s = \lambda/4L$ is a dimensionless factor and called as quasi-wavelength (QW). Equation 4 looks like the typical formula of a Michelson interferometer. Since the integral part $N$ is usually unknown for phase measurement, the unambiguous range of $(n-1)$ in Equ.4 equals to $\lambda_s$. To extend the range in Equ.4, following with the route of synthetic wavelength theory in optical interferometry[14-15], we choose two vacuum tubes with lengths of $L_1$ and $L_2$ to measure the refractive index simultaneously. Two formulas can be obtained according to Equ.4

$$\left.\begin{array}{l} n-1=\lambda_{s1}\left(N_1+\varepsilon_1\right) \\ n-1=\lambda_{s2}\left(N_2+\varepsilon_2\right) \end{array}\right\}, \quad (5)$$

where $\lambda_{s1} = \lambda/4L_1$ and $\lambda_{s2} = \lambda/4L_2$ ($L_1 > L_2$). The subtraction between two formulas in Equ.5, we obtain the following formula

$$n-1=\Lambda_s\left(\Delta N+\Delta\varepsilon\right), \quad (6)$$

where $\Delta N$ equals to $N_1 - N_2$, $\Delta\varepsilon$ equals to $= \varepsilon_1 - \varepsilon_2$, and $\Lambda_s = \lambda/4\left|L_1 - L_2\right|$ is the synthetic factor calculated from two QWs and called as quasi-synthetic wavelength (QSW). Comparing with Equ.4 with Equ.6, it can be apparently concluded that the unambiguous range in Equ.4 is extended from $\lambda_{si}$ ($i = 1,2$) to $\Lambda_s$. Supposing $\lambda$ equals to 633nm, $L_1$ equals to 165mm and $L_2$ equals to 160mm, the unambiguous range for $\lambda_{s1}$ and $\lambda_{s2}$ is $0.96 \times 10^{-6}$ and $0.99 \times 10^{-6}$, respectively. But that of $\Lambda_s$ is $31.5 \times 10^{-6}$, which is magnified by nearly 32 times.

In the same way shown above, a QSW chain can be organized by choosing a group of tubes with proper lengths. For consistency, we call the QWs as zero-order QSW, and higher order QSWs can be generated according to Equ.6. A case of QSW chain organized by three tubes (with lengths of $L_1$, $L_2$ and $L_3$, supposing $L_1 > L_2 > L_3$) has been shown in Fig.2.

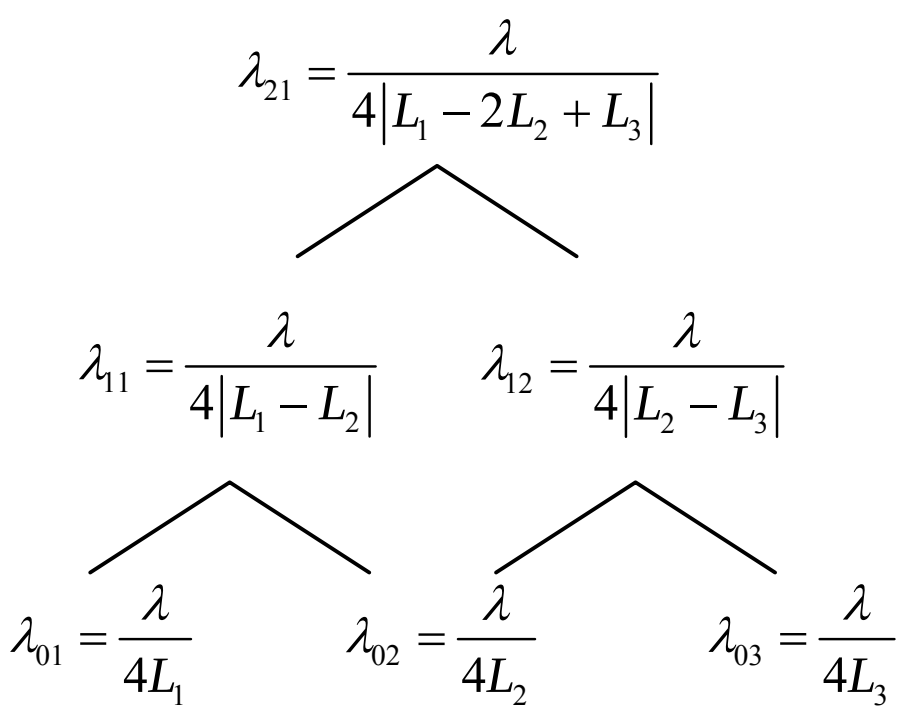

**Fig.2** A case of QSW chain organized by three vacuum tubes. The symbol $\lambda_{ij}$ represents the QSWs, in which $i$ indicates its order and $j$ indicates its number

If the highest order QSW shown in Fig.2 is larger than $(n-1)$, the refractive index of measured gas can be determined absolutely by this QSW chain in the following way. In the first place, for the highest order QSW $\lambda_{21}$, the refractive index can be deduced roughly by the fractional part of the phase shift in this order. In the second place, this rough value of refractive index is served as an input data to estimate the integral part $N$ in the near lower order QSW $\lambda_{11}$ and $\lambda_{12}$. The estimating formula can be written as

$$N_{j-1} = \mathrm{INT}\left( \frac{(n-1)_j}{\lambda_{j-1}} - \varepsilon_{j-1} + \frac{1}{2} \right), \ j = 1, 2, \ldots, \quad (7)$$

where INT means rounding the element in the bracket to the nearest integer that no more than itself, and $j$ indicates the order of QSWs. And the refractive index can be determined by combining $N_1$ and fractional part $\varepsilon_1$. Following the same manner above, the refractive index can be determined by zero-order QSWs finally. In practice, the fractional phases ( $\varepsilon_{01}$, $\varepsilon_{02}$ and $\varepsilon_{03}$ )of zero-order QSWs can be measured directly when each tube is in the optical path, and that of higher order QSWs are calculated according to Equ.6.

### *2.2 Length Design of Vacuum Tube*

The length design of vacuum tube is of great importance for a gas refractometer, and three issues, namely, the refractive index of measured gas, expected accuracy, and transition restriction, should be concerned properly before constructing an apparatus. At first, the highest order QSW should be larger than the refractive index of gas we want to measure, so that the refractometer can be performed in an absolute way. Moreover, the expected accuracy is relevant to the longest vacuum tubes. In the end, a transition restriction should be satisfied if the measured value by a higher-order QSW can be served as an input data to estimate the integral part of near lower-order QSW without ambiguity. This transition restriction can be expressed as

$$u(n-1)_j < \frac{1}{2}\lambda_j - u(n-1)_{j-1}, \quad (8)$$

where $u$ represents the measurement uncertainty of the refractivity in the j-order QSW.

According to aforesaid three issues, we demonstrate a design example here. Three tubes with lengths of $L_1$, $L_2$ and $L_3$ ( supposing $L_1 > L_2 > L_3$ ) are used for constructing the QSW chain. The measurement range is $R$, and the measurement uncertainty of tube length and phase shift is $u(L)$ and $u(p)$, respectively. The uncertainty of the laser's wavelength is usually so small that can be ignored here. The refractive index determined by each order QSW has an uncertainty of $u(RI_j)$, where $j$ indicates the QSW's order. If the expected measurement uncertainty of refractive index is $u(RI_m)$, then the restrictions of tubes' lengths can be summarized as

$$
\left.\begin{array}{l}
\dfrac{\lambda}{4(L_1-2L_2+L_3)} > R \\
u\left(RI_2\right) < \dfrac{1}{2}\cdot\dfrac{\lambda}{4(L_1-L_2)} - u\left(RI_1\right) \\
u\left(RI_1\right) < \dfrac{1}{2}\cdot\dfrac{\lambda}{4L_1} - u\left(RI_0\right) \\
u\left(RI_0\right) < u\left(RI_m\right)
\end{array}\right\}. \quad (9)
$$

According to Equ.9, a group of vacuum tubes with optimized lengths can be selected. To discuss it in details, we suppose the measured gas is air, whose refractive index will be no more than 1.0003 in normal condition. We then estimate that $u(L)$ and $u(p)$ equals to 0.002 mm and 0.005, respectively. If the expected measurement uncertainty of refractive index at $\lambda$ =633 nm is a few parts in $10^8$, the result of length design for three tubes is shown in Fig.3.

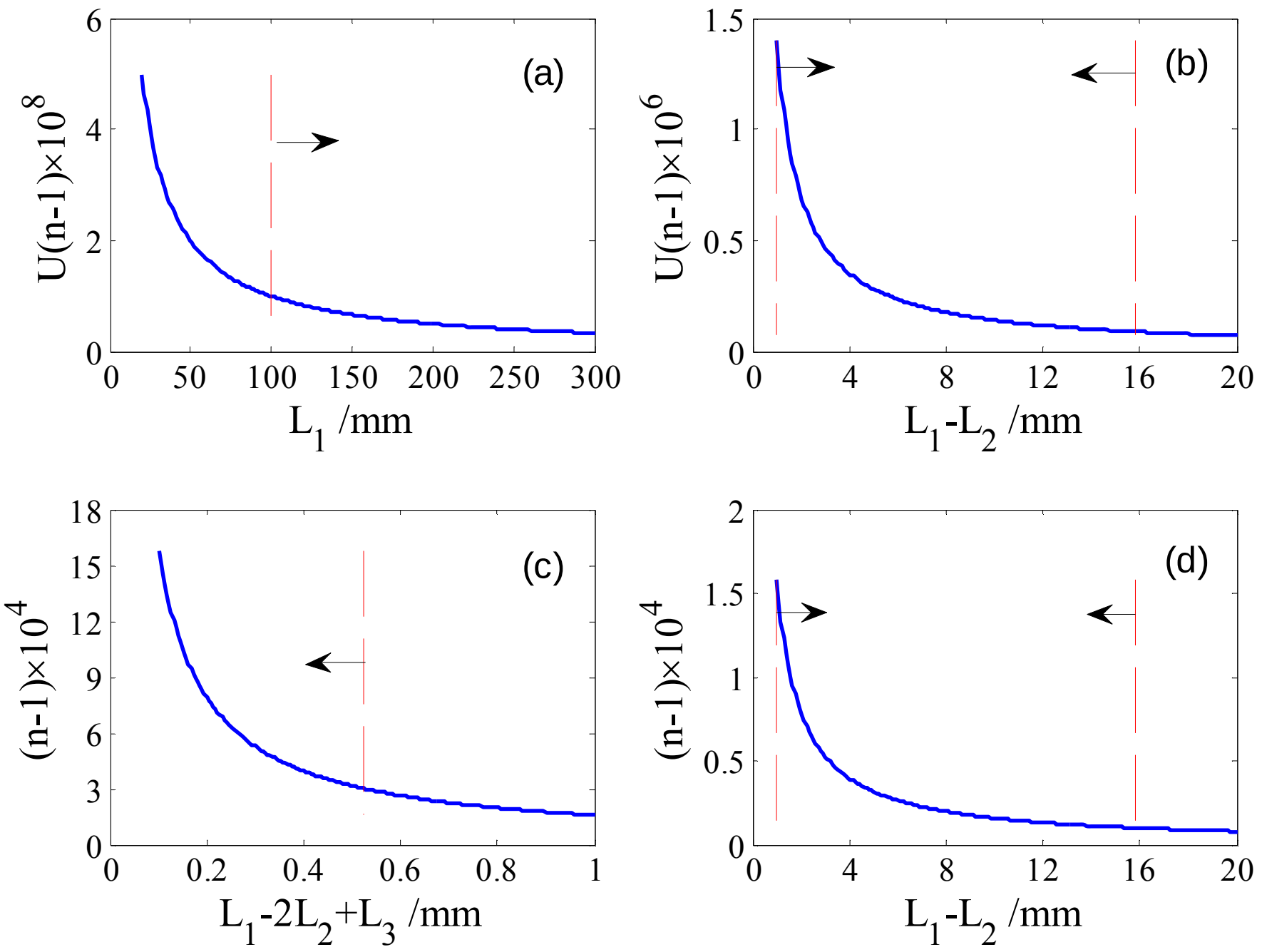

**Fig.3** An example for length design of vacuum tube. The dish line and arrow indicate the feasible range for length selection. (a) Restriction of the length L1 caused by the expected measurement uncertainty. (b) Restriction of the length L1-L2 caused by the measurement uncertainty by means of first-order QSW. (c) Restriction of the length L1-2L2+L3 caused by the measurement range. (d) Restriction of the length L1-L2 caused by the measurement range by means of first-order QSW.

The measurement range and uncertainty of a gas refractometer is variable by the combination of vacuum tubes with proper lengths according to Fig.3. For example, it is indicated that the expected measurement uncertainty is about 1E-8 when the length of L1 is about 100 mm, and it can be improved to 5E-9 when L1 is about 200 mm. Moreover, the expected range of (n-1) is about 3E-4 if the value of (L1-2L2+L3) equals to about 0.52 mm, and it also can be extended by four times when (L1-2L2+L3) equals to about 0.13 mm. The flexible measurement range from 1 to 1.0012 can cover almost all of the routine gases [18]. It should be noted that, however, the larger range requires the more rigorous transition restriction, which means more vacuum tubes and higher QSW order may be needed.

## 3 Experiments and Results

### *3.1 Measurement Apparatus*

The experiment setup of gas refractometer is shown in Fig.4. The vacuum tubes are made of BK7 glass, and its diameter is 18 mm with wall thickness of 5 mm. The end mirror has a diameter of 42 mm and thickness of 5mm, and the transmittance is better than 99.8% after a coating of anti-reflection film. The end mirrors stick to the end faces of the vacuum tube in parallel by epoxy resin, and the inner space of the tube is pumped down to $10^{-2}$ Pa before sealed. According to the QSW method presented in Sec.2, a group of three vacuum tubes with nominal length of 165 mm, 158 mm and 151.5 mm is selected for construction of the QSW chain. The final lengths after optical fabrication are 164.5864 mm, 157.8515 mm and 151.5166 mm. Therefore, the expected measurement range and uncertainty of the refractive index is about 3.95E-4 and 6E-9, respectively, according to the estimation in Sec.2.2. Three vacuum tubes are mounted in parallel on a motorized 1D displacement stage, which can transport the tubes into the optical path in length sequence. The phase data is recorded by the PM when each vacuum tube is in the optical path, and the refractive index of measured gas can be determined by means of QSW method. In experiment, an additional phase data is also recorded when there is no tube in the path. It is a phase delay caused by other optical elements except the tubes, and should be subtracted from the measured phases of vacuum tubes before further calculation. A dual-frequency laser head (Agilent 5517B) is used as the light source, the beam emitted from which consists of two linear polarized frequency components with orthogonal direction. The vacuum wavelength is 632.991372 nm with an uncertainty better than $\pm$1E-7(3$\sigma$), and the split frequency of two components is 2.2 MHz. The bandwidth of the photo detectors D1 and D2 (Agilent 10780F) is about 7 MHz. The PM (Pretios PT-1313F) has an accuracy of 3.6 degree.

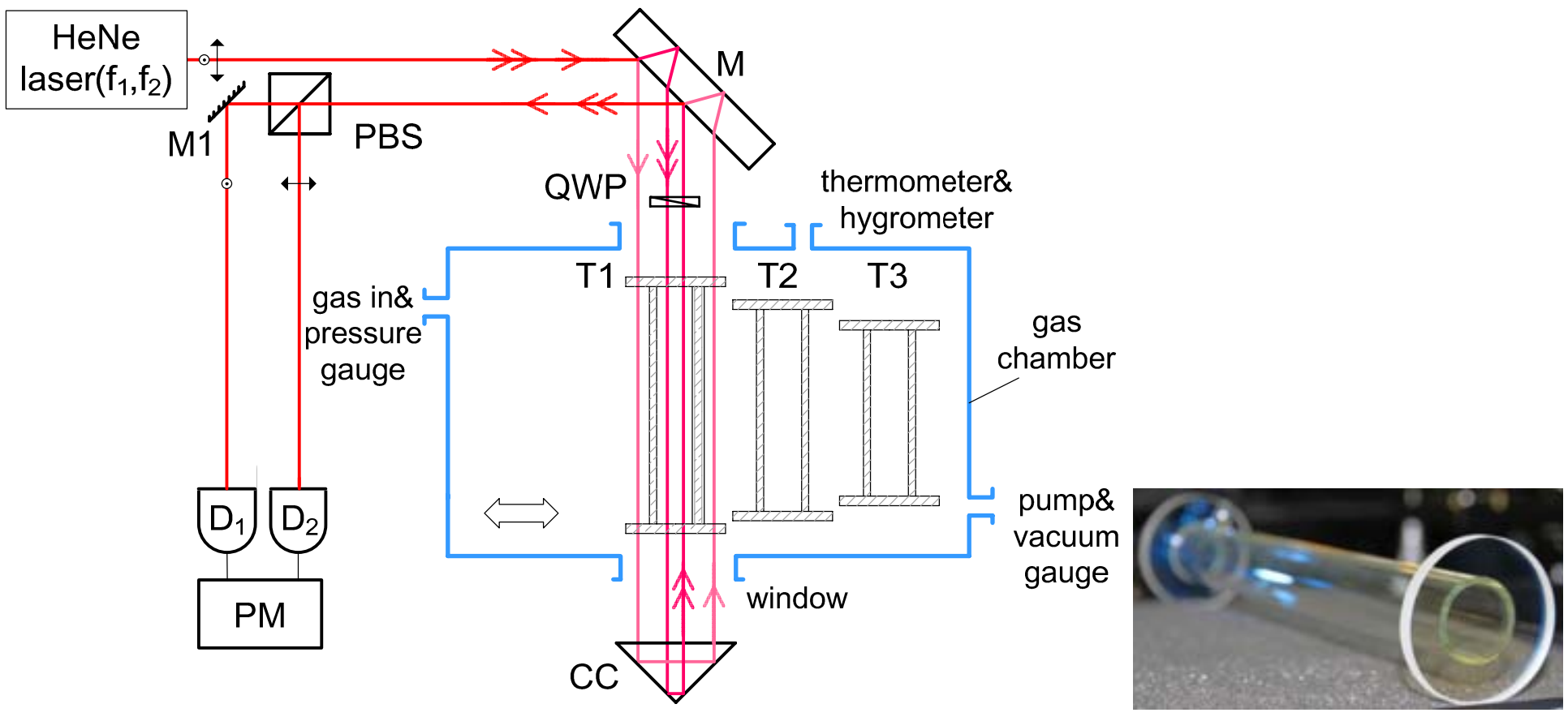

**Fig.4** Optical setup of gas refractometer. HeNe laser, laser with two orthogonally linear polarized frequency components f1 and f2; M, plane beam splitter; QWP, quarter wave plate; T1, T2 and T3, vacuum tubes; CC, corner cube; PBS, polarized beam splitter; M1, reflector; D1, D2, photodetectors; PM, phase meter. The photograph of a vacuum tube is also shown.

The optical system is placed on a breadboard for vibration isolation, and the vacuum tubes are mounted in a gas chamber. There are two quartz glass windows on the chamber for light transmission. The measured gas can be filled in or pumped out through two pipe equipped on the chamber. The environment parameters inside the chamber are also monitored by several sensors, namely, the thermometer (Fluke 1523), pressure gauge (Setra Model470) and hygrometer (Rotronic HP23). In addition, the chamber should be pumped to vacuum to avoid gas contamination before each measured gas is charged, and the residual gas is monitored by a vacuum gauge (Inficon CDG025D-X3). All the sensors are carefully calibrated by the National Institute of Metrology of China, and the calibrated results are summarized in Table.1.

Table.1 Calibrated results of thermometer, pressure gauge, hygrometer and vacuum gauge

| Sensor | Calibrated range | Calibrated uncertainty |
|---|---|---|
| Temperature (℃) | 14~26 | 0.01 |
| Pressure (kPa) | 80~110 | 0.014 |
| Humidity (RH%) | 20~60 | <2 |
| Vacuum (Pa) | 0~1000 | 2% of reading |

### *3.2 Experiment Results*

There already have been some precision data measured by previous methods for air and nitrogen gas, which can be used by us as reference. Therefore, to verify the performance of the gas refractometer, dry air and nitrogen gas (purity >99.9999%) are measured in our experiments. The chamber is filled in and pumped out the measured gas for several times to avoid the contamination before measurement. The chamber is then filled in the measured gas smoothly. The gas pressure in the chamber is controlled manually by a valve to specific values, roughly from 80 kPa to 105 kPa with a step of 5 kPa. Five repeated measurements are performed at each pressure value after the gas is balanced. The measurement results are compared with the reference data deduced by the empirical formula. For dry air, the reference data is calculated by the modified Edlén equation deduced by Birch[1-2]. And for nitrogen gas, since the formulas deduced by Peck[16] and Zhang[17] are consistent with each other at 1.1E-8 level at 633 nm, the latter is adopted. The experiment results are shown in Fig.5 and 6.

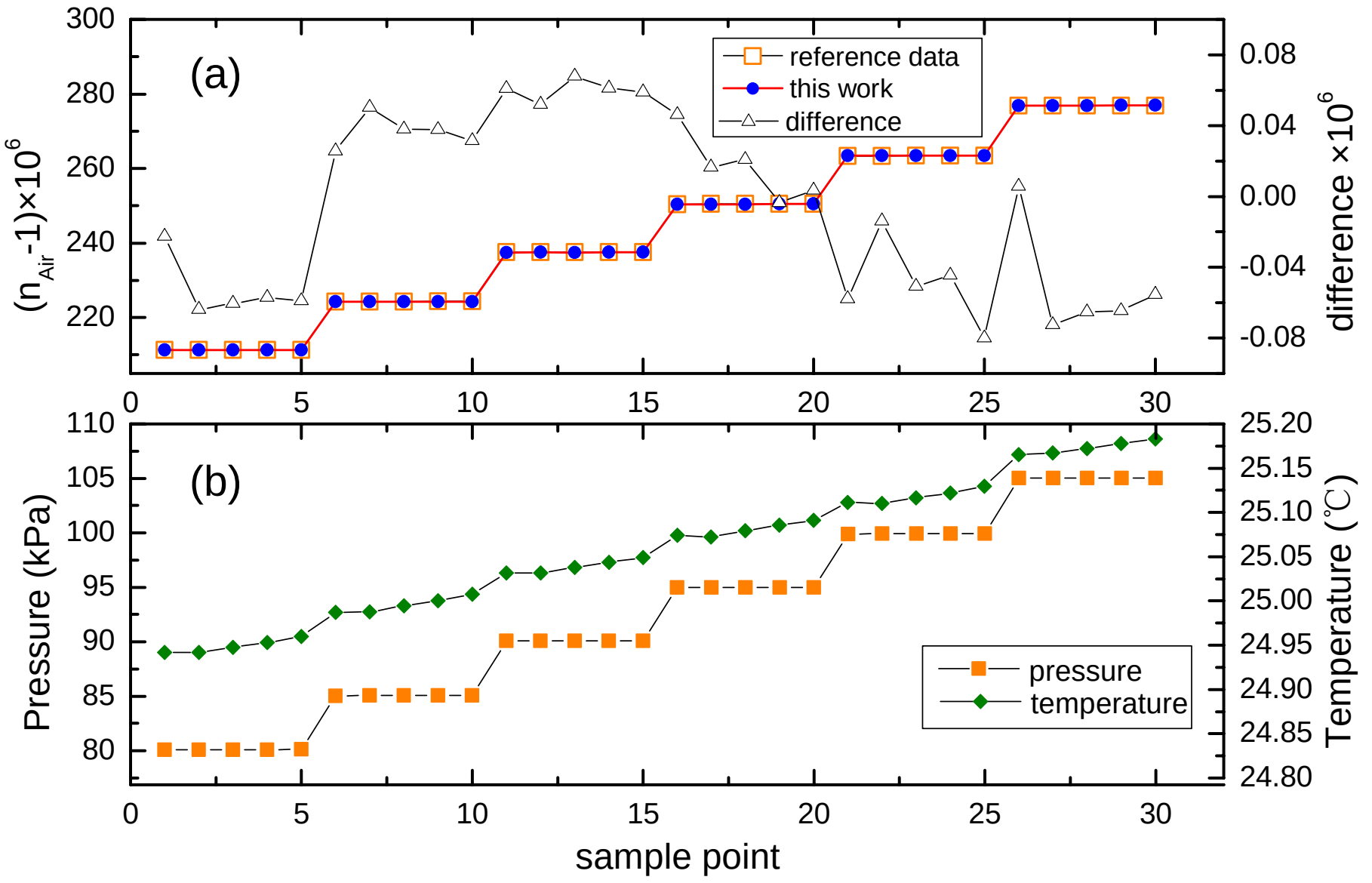

**Fig.5** Refractive index of dry air at 633 nm. (a) Experiment and reference data of dry air. (b) Pressure and temperature data in the measurement

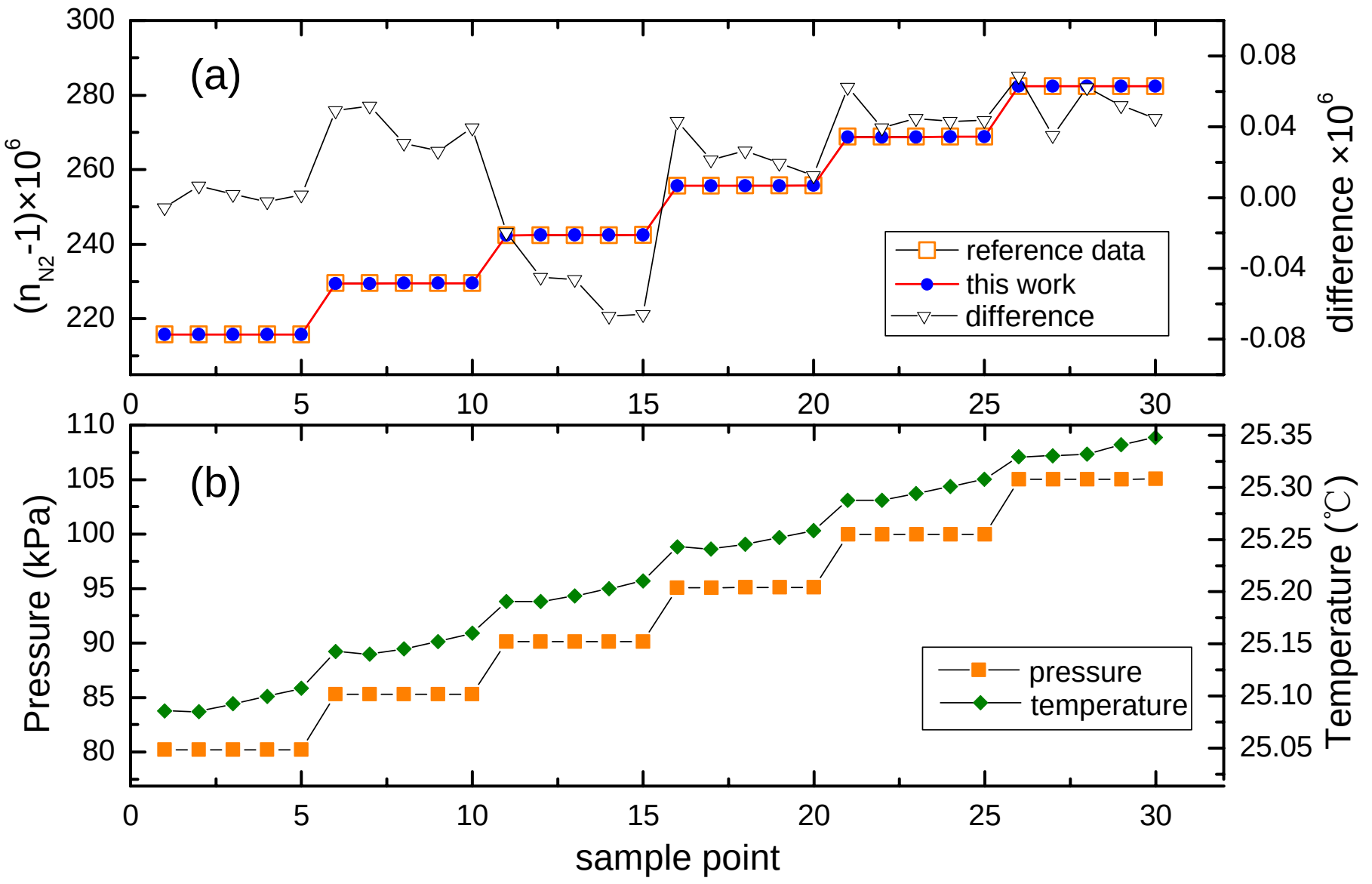

**Fig.6** Refractive index of nitrogen gas at 633 nm. (a) Experiment and reference data of nitrogen gas. (b) Pressure and temperature data in the measurement

Moreover, a long term measurement of ambient air has also been performed by this refractometer to test its stability. It takes about 16 h for this continuous measurement, and the result is shown in Fig.7.

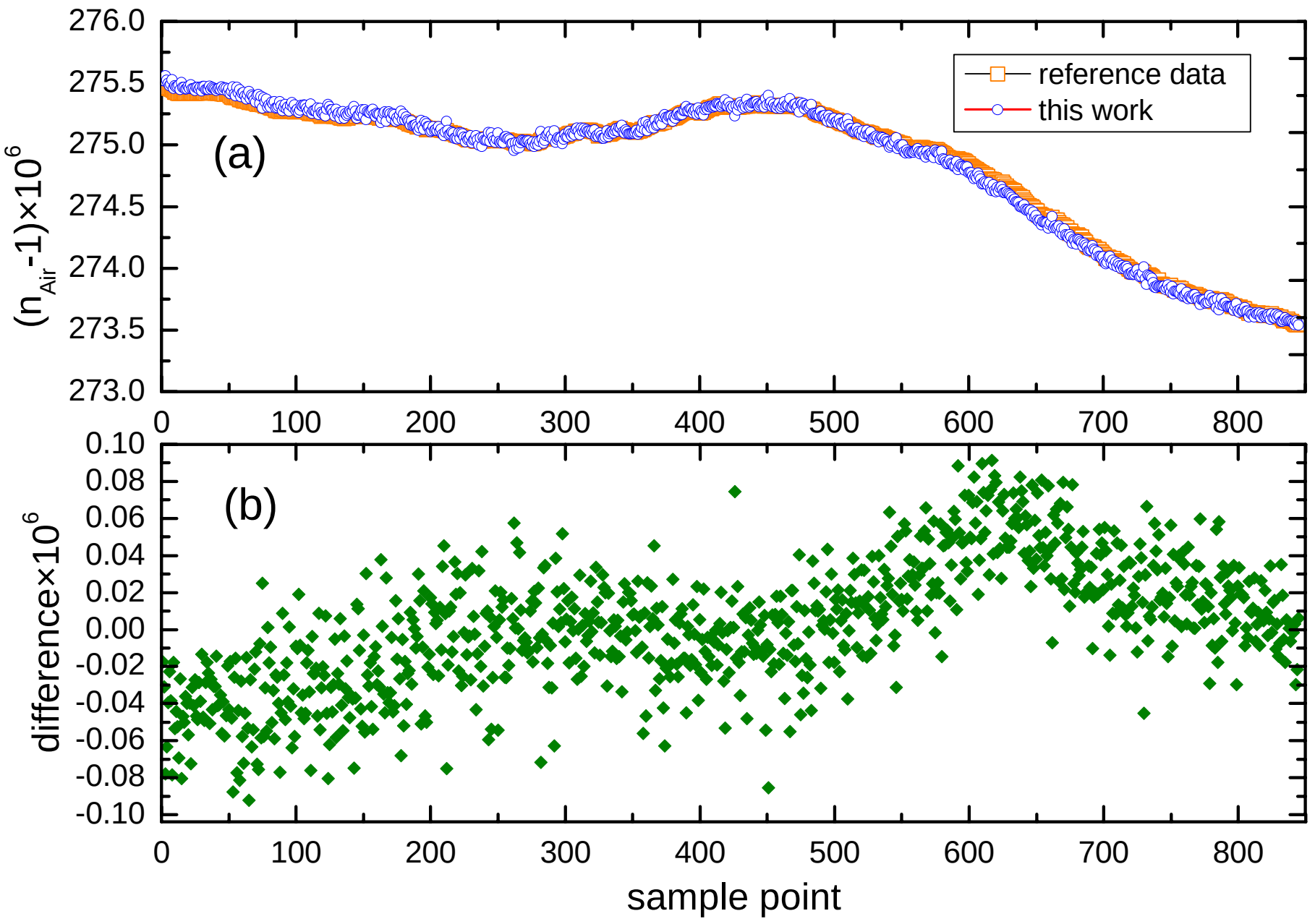

**Fig.7** Long term measurement of refractive index of ambient gas at 633 nm for about 16 h. (a) Experiment and reference data of ambient air. (b) Difference between experiment and reference data.

Within the measurement range from 1.00022 to 1.00028, the consistence between our refractometer and previous methods is within the range of 8E-8 for the measurement of dry air and nitrogen gas according to Fig.5&6. And the long term measurement of the refractive index of ambient gas in Fig.7 indicates a good stability of the performance. It should be stated that, however, besides the measurement data, the reference data partly accounts for the above discrepancy. Since the reference data are calculated by the empirical equation with environmental parameters (such as temperature, pressure and humidity), and the calibrated uncertainty of these sensors will contribute an uncertainty of about 5E-8 to the reference data.

### *3.3 Uncertainty Evaluation*

According to Equ.4 and the principle of QSW method, the measurement uncertainty of this refractometer is ultimately determined by that of zero-order QSW, which can be described as

$$\frac{u(n-1)}{n-1}=\left[\left(\frac{u(\varepsilon_{01})}{N_{01}+\varepsilon_{01}}\right)^2+\left(\frac{u(L_1)}{L_1}\right)^2+\left(\frac{u(\lambda)}{\lambda}\right)^2\right]^{1/2}, (10)$$

where $N_{01}$ and $\varepsilon_{01}$ is the integral and fractional part of phase shift when measuring the refractive index by zero-order QSW $\lambda_{01}$.

A. Wavelength error: the relative uncertainty of the laser head is better than 1E-7, and its contributions to the measurement is about a few part in $10^{11}$, which can be ignored.
B. Phase error: two sources will account for this. The first is the measurement uncertainty of the phase meter, which is about 3.6 degree and has a contribution of 1E-8 to the measurement of refractive index. The second is the additional phase caused by the optical elements in the system. For example, the azimuth error of the quarter wave plate's fast axis or the delayed phase deviating from exact 90 degree will introduce crosstalk between two linear polarized components, and additional phase will mixed with the target one. Moreover, the extinction ratio and adjustment of the polarized beam splitter will affect the phase in the similar manner. However, since these effects are almost the same for three vacuum tubes, the additional phase measurement when there is no tube in the optical path (see Sec.3.1) can eliminate them effectively. And the residual phase error will have a contribution of about 1E-8.
C. Length error of vacuum tube: The length of vacuum tube is measured by the coordinate measuring machine with an uncertainty of less than 0.01 mm. In addition, the temperature has effect on the tube's length. The thermal expansion of BK7 glass is 7.1E-6 /K, and for the longest vacuum tube, the temperature fluctuation within 10 K around 295.15 K will introduce an error of 0.0117 mm. In addition, if the vacuum tube is not in parallel with the optical path, the real path passed by the beam will be larger than the geometric length of the tube. After careful adjustment, the tilt angle can be confined within 0.2 degree, and this contribution to the measurement is about 1E-9. In short, the contribution of the length error is about 2.8E-8 for the measurement of refractive index.
D. Residual gas in vacuum tube: If there is residual gas inside the vacuum tube, a systematic error will be introduced to the final result of the gas refractometer. The ambient air inside the vacuum tube has been first exhausted below 0.02 Pa by molecular vacuum pump, and then the glass pipe connected to the vacuum tube is melted and sealed. The measurement error caused by residual gas with 0.02 Pa will be no more than 6E-11.

Overall, the combined uncertainty of the gas refractometer is about 3.1E-8.

**4 Conclusions**

We have reported a gas refractometer capable of measuring the refractive index of gas as large as 1.000395 at 633 nm absolutely with an uncertainty of about 3.1E-8 based on the QSW method. The extended measurement range of the refractometer is realized by a group of home-made vacuum tubes according to the QSW method. In addition, there is no gas-filling or pumping process during the measurement, so that a complete measurement costs only 70 seconds. The principle of QSW method and the design of refractometer are presented in detail. The performance of this refractometer has been verified by measuring the refractive index of dry air, nitrogen gas and ambient air. The simple configuration makes it compact and easy to handle. Comparing with the previous refractometers, the refractometer presented here has integrated virtues of large unambiguous range, fast speed, high accuracy, and a simple instrumentation design. The measurement range may be extended further to several times of current one according to the QSW method. In addition, the measurement time may be reduced to within 10 seconds by a new arrangement of optical path which can measure the phases of all the vacuum tubes simultaneously. All of these possible improvements are under consideration for future work.

**Acknowledgement**: The authors are grateful to the partly support of the National Science and Technology Major Project of China.

**References**

1. K. P. Birch and M. J. Downs, “An updated Edlén equation for the refractive index of air,” Metrologia 30, 155-162(1993).
2. K. P. Birch and M. J. Downs, “Correction to the updated Edlén equation for the refractive index of air,” Metrologia 31, 315-316(1994).
3. P. E. Ciddor, “Refractive index of air: new equations for the visible and near infrared,” Appl. Opt. 35, 1566-1573(1996).
4. G. Bönsch and E. Potulski, “Measurement of the refractive index of air and comparison with modified Edlén’s formulae,” Metrologia 35, 133-139(1998).
5. M Ishige, M Aketagawa, T B Quoc and Y Hoshino, “Measurement of air-refractive-index fluctuation from frequency change using a phase modulation homodyne interferometer and an external cavity laser diode,” Meas. Sci. Technol. 20, 084019(2009)
6. K. P. Birch, “Precise determination of refractometric parameters for atmospheric gases,” J. Opt. Soc. Am. A 8, 647-651(1991).
7. Q. H. Chen, H. F. Luo, S. M. Wang, and F. Wang, “Measurement of air refractive index based on surface plasmon resonance and phase detection,” Opt. Lett. 37, 2916-2918(2011).
8. L. J. Zeng, I. Fujima, A. Hirai, H. Matsumoto, and S. Iwasaki, “A two-color heterodyne interferometer for measuring the refractive index of air using an optical diffraction grating,” Opt. Comm. 203, 243-247(2002).
9. J Terrien, “An air refractometer for interference length metrology,” Metrologia 1, 80-83(1965).

10. P. Egan and J. A. Stone, "Absolute refractometry of dry gas to ±3 parts in 109," Appl. Opt. 50, 3076-3086(2011).
11. J. Zhang, Z. H. Lu, and L. J. Wang, "Precision measurement of the refractive index of air with frequency combs," Opt. Lett. 30, 3314-3316(2005).
12. J. Zhang, Z. H. Lu, B. Menegozzi and L. J. Wang, "Application of frequency combs in the measurement of the refractive index of air," Rev. Sci. Instrum. 77, 083104(2006).
13. G. P. Xie, Y. J. Wang, L. J. Reiter, and J. C. Tsai, "High-accuracy absolute measurement of the refractive index of air," Opt. Eng. 43, 950-953(2004).
14. P. J. Groot, "Extending the unambiguous range of two-color interferometers," Appl. Opt. 33, 5948-5953(1994).
15. S. H. Lu and C. C. Lee, "Measuring large step heights by variable synthetic wavelength interferometry," Meas. Sci. Technol. 13, 1382-1387(2002).
16. E. R. Peck and B. N. Khanna, "Dispersion of nitrogen," J. Opt. Soc. Am. 56, 1059-1063(1966).
17. J. Zhang, Z. H. Lu, and L. J. Wang, "Precision refractive index measurements of air, N2, O2, and CO2 with a frequency comb," Appl. Opt. 47, 3143-3150(2008).
18. M J Weber Handbook of optical materials (CRC Press, London, 2003) p.445-460